\pdfoutput=1
\documentclass[aps,amsfonts,amsmath,prd,preprint,nofootinbib]{revtex4}

\usepackage{natbib} 
\usepackage[utf8]{inputenc}
\usepackage{hyperref}
\usepackage{xcolor}
\usepackage{graphicx}
\graphicspath{{./img/}}
\usepackage{amsmath,amssymb,amsbsy,amstext,amsthm}
\usepackage{mathtools}
\usepackage{amsfonts}
\usepackage{comment}
\usepackage{bm}
\usepackage{tabularx}
\usepackage{array,collcell}

\usepackage{url}                                 %Supports url commands

\def\be{\begin{equation}}
\def\ee{\end{equation}}
\newcommand{\dd}{\mathrm{d}\baselineskip0.5ex}
\def\Ms{{M_{\odot}}}

\def\tdec{{t_{\rm dec}}} 

\def\teq{{t_{\rm eq}}}

\def\tobs{{t_{\rm obs}}}
\def\tff{{t_{\rm ff}}}

\def\xM{{x_{\rm max}}}
\def\xm{{x_{\rm min}}}

\newcommand{\eqdot}{\, .}
\newcommand{\eqcom}{\, , \quad}

\def\bea{\begin{eqnarray}}
\def\eea{\end{eqnarray}}
\def\be{\begin{equation}}
\def\ee{\end{equation}}
\def\ba{\begin{array}}
\def\ea{\end{array}}

\newcommand{\beq}{\begin{equation}}
\newcommand{\eeq}{\end{equation}}

\usepackage{graphicx,subfigure}

\usepackage{amsmath}

\def\bea{\begin{eqnarray}}
\def\eea{\end{eqnarray}}

\definecolor{DarkBlue}{rgb}{0,0,0.7} 

\definecolor{DarkRed}{rgb}{0.85,0,0} 

\begin{document}

\title{Constraining changes in the merger history of (P)BH binaries 
with the stochastic gravitational wave background }
%of past mergers}

\author{Vicente Atal${}^{a}$, Jose J. Blanco-Pillado${}^{a,b}$, Albert Sanglas${}^{c}$ and Nikolaos Triantafyllou${}^{c}$}
\affiliation{$^a$ Department of Theoretical Physics, University of the Basque Country, Bilbao, Spain}
\affiliation{$^b$ IKERBASQUE, Basque Foundation for Science, 48011, Bilbao, Spain}
\affiliation{$^c$ Departament de F\'{i}sica Qu\`{a}ntica i Astrof\'{i}sica, i Institut de Ci\`{e}ncies del Cosmos, Universitat de Barcelona, Mart\'{i} Franqu\`{e}s 1, 08028 Barcelona, Spain.}

%\fntext[fn1]{vicente.atal@ehu.eus}
%\fntext[fn2]{josejuan.blanco@ehu.eus}
%\fntext[fn3]{asanglas@icc.ub.edu}
%\fntext[fn4]{nitriant@icc.ub.edu}

\begin{abstract}
Black holes binaries coming from a distribution of primordial black holes might exhibit a large merger rate up to large redshifts. Using a phenomenological model for the merger rate, we show that  changes in its slope, up to redshifts $z\sim 4$, are constrained by current limits on the amplitude of the stochastic gravitational wave background from LIGO/Virgo -O3 run.  This shows that the stochastic background constrains the merger rate for redshifts larger than the single event horizon of detection ($z\simeq 1$ for the same detector). Moreover, we show that for steep merger rates the shape of the stochastic gravitational wave signal at intermediate frequencies differs from the usual $2/3$ IR scaling.
 We discuss the implications of our model for future experiments in a wide range of frequencies, as the design LIGO/Virgo array, Einstein Telescope, LISA and PTA. Additionally, we show that i) the stochastic background and the present merger rate provide equally constraining bounds on the abundance of  PBHs arising from a Gaussian distribution of inhomogeneities and ii) Degeneracies at the level of the stochastic background (i.e. different merger histories leading to a similar stochastic background) can be broken by considering the complementary direct observation of the merger rate from number counts by future detectors such as the Einstein Telescope. 
%We finally discuss the prospects of Einstein telescope for achieving this goal. 
\end{abstract}

\maketitle

\section{Introduction}
\noindent
The GWTC-3 catalog of LIGO/Virgo reports the direct detection of 63 binary black hole (BBH) mergers from redshifts $z\simeq0.1$ up to $z\simeq0.8$ \cite{LIGOScientific:2021psn}. 
With these events it is possible to constrain the merger rate density of BH binaries as a function of redshift. In particular, for a merger rate density following a power law distribution given by
\be\label{eq:canonical_model}
\frac{dN_{\rm merg}}{dtdV}\equiv R(z)= R_0(1+z)^\alpha,
\ee
the factor $\alpha$ is found to be $2.7^{+1.8}_{-1.9}$ \cite{LIGOScientific:2021psn} for a present merger rate $R_{\rm 0}=9-35 ~\rm {Gpc^{-3}yr^{-1}}$. These constraints are evidently only sensitive up to the largest redshift at which the merger of a certain binary might be detectable. This is the so-called horizon distance, and for BBH of $\sim 30 M_{\odot}$, $z_{\rm hor}\simeq 1$ for LIGO/Virgo. An important question is whether any relevant information at redshifts larger than $z_{\rm hor}$ can be inferred from GW observations.  

The answer is positive, since, apart from the direct detections of BBHs mergers, gravitational wave detectors are sensitive to the stochastic background sourced from past mergers that are not necessarily individually distinguishable. The stochastic background is proportional to the integrated merger rate over time, and therefore it might be sensitive to the evolution of the merger rate up to redshifts larger than those inferred from single detections. 

At the moment such background has not been detected. This absence has been primarily used to constrain the BBH abundance \cite{Rosado:2011kv,Wu:2011ac,Mandic:2016lcn,Wang:2016ana,Clesse:2016ajp,Raidal:2017mfl,Raidal:2018bbj,Mukherjee:2021ags,Mukherjee:2021itf,Callister:2020arv,Inayoshi:2021atf}.  Here we will mostly study how this lack of observation can constrain the merger rate. This direction has been taken by some recent works \cite{Callister:2020arv,Mukherjee:2021itf} (see also \cite{Bavera:2021wmw}), including the analysis of LIGO/Virgo collaboration \cite{KAGRA:2021kbb}.  Here we will complement these results with a more phenomenological approach that will allow us to determine more precisely the maximum redshift that plays a role in determining the stochastic background (and thus the maximum redshift at which the merger rate might be constrained from (non) observations of the stochastic background), as well as  concentrating on the influence of the merger rate on the resulting shape for the stochastic background. We will also argue for an extension of the parameter space studied in these works. 

In general, it is assumed that the binary merger rate has either the form of  Eq. \eqref{eq:canonical_model} up to an arbitrary redshift, or functions that are given by Eq. \eqref{eq:canonical_model} up to a certain redshift and later decay. These two families of models roughly correspond to primordial and stellar black holes respectively. Here we will concentrate in the former case (although not precluding the possibility of a mixed population). Primordial black holes (PBHs) are interesting given that they might not only constitute part or the totality of the dark matter (DM) \cite{Sasaki:2016jop,Bird:2016dcv,Clesse:2016vqa}, but could also inherit valuable information of a primordial era of evolution of our Universe (for recent reviews on PBHs and DM, see \cite{Sasaki:2018dmp,Green:2020jor}).
For PBHs evolving from an initial Poissonian distribution,  - as it happens if the fluctuations leading to the PBHs are Gaussian \cite{Ali-Haimoud:2018dau,Desjacques:2018wuu,Ballesteros:2018swv}- it is actually possible to determine the exponent $\textstyle{\alpha}$ of the merger rate density in Eq. \eqref{eq:canonical_model}. For small abundances, $f_{\rm PBH}\lesssim 0.01$, and small redshifts, $z<1$, we have that $\alpha= 1.1$. For larger redshifts, $\alpha=1.4$ (see section \ref{sec:1}). The behaviour of the merger rate with redshift in the case of both large abundances and clustered distributions remains however unclear\footnote{Note that N-body effects might also be important for small abundances \cite{Garriga:2019vqu}}. For large abundances binaries can not be treated as isolated objects and N-body effects have to be considered in order to determine the merger rate evolution \cite{Raidal:2018bbj,Jedamzik:2020omx}. This is also a problem for clustered distributions, with the additional complication that the initial distribution could also depend on an enlarged set of parameters (given by -in principle- independent N-point correlations functions).
%
%More precisely, while a Poissonian distribution depends on a single parameter, which in the case of PBHs can be taken as its abundance relative to dark matter, there are in principle infinite parameters for describing a non-Poissonian distribution, given by independent N-point correlations functions. While we expect that these independent N-point correlation function depend on a few parameters of the parent theory of the fluctuations, there is no known model independent relation that allows us to span all the posibles distributions of clustered PBHs in a simple way \footnote{In some works \cite{DeLuca:2021hde} it has been argued that the merger rate obtained following the description of a Poissonian distribution plus a locally enhanced number density of PBHs is an upper bound on the true merger rate (the "actual" merger rate being smaller given the disruption of binaries by N-body dynamics). This is not necessarily true, since the calculations of the merger rate following the description of the Poissonian distribution assumes the separability condition of the correlation functions, which is not necessarily satisfied in clustered distributions \cite{Ballesteros:2018swv,Atal:2020igj}.}.
%
Despite these dificulties, some interesting results have been obtained in determining the merger rate for some families of clustered distributions, with the use of numerical simulations  \cite{Trashorras:2020mwn,Antonini:2020xnd}. For example, BBHs merging inside globular clusters can exhibit $\alpha=2.3^{+1.3}_{-1.0}$ \cite{Antonini:2020xnd}. Other studies focusing on non-Gaussian initial distributions of PBHs -but having neglected the effects of N-body disruption of binaries- have shown that the merger rate of PBH binaries can have a very complex dependence on redshift, in particular, with an interpolation of various slopes at intermediate redshifts \cite{Atal:2020igj}. 

These results are a clear motivation for exploring a larger set of BBHs merger histories. Given that there is still a large uncertainty in the class of allowed merger histories, we adopt a phenomenological approach and consider the consequences of a change in the slope of the merger rate at some particular redshift.
The primary goal of this paper is to determine how the non-detection of a stochastic gravitational wave by the LIGO/Virgo collaboration constrains the changes in the merger rate. As we will see, this lack of detection allows us to constrain the changes in the slope at redshift larger than the horizon redshift $z_{h}$. We will also explore how future experiments could tighten these bounds. We will test the future LIGO/Virgo observations and, in the same frequency band, Einstein Telescope \cite{Punturo:2010zz}. We will briefly comment on the experiments at smaller frequencies such as LISA \cite{LISA:2017pwj} and PTA observatories (in particular NANOGrav \cite{NANOGrav:2020bcs} and SKA \cite{Janssen:2014dka}).

The paper is organised as follows. In Section \ref{sec:1} we first show how a wide variety of binary merger histories can be obtained in clustered distribution of BHs, and present a model for the merger rate that takes into account such variety. In Section \ref{sec:2}  we analyse the resulting stochastic background generated by the merging population of BBHs with a monochromatic mass distribution. In Section \ref{sec:3} we introduce the expected signal-to-noise ratio (SNR) for various detectors and the so-called abundance criterion. Both serve as useful constraints of the model's parameter space. In Section \ref{sec:4} we present the resulting constraints coming from the non-observation of the stochastic signal, and calculate the minimum SNR ratio that should be detected in forthcoming experiments if the model under consideration accounts for BBHs. In \ref{sec:ET} we show how a direct inference of the merger rate with ET can help us breaking the degeneracies existing at the level of the stochastic background. The results are discussed in Section \ref{sec:conclusions}. In Appendix \ref{sec:appendix} we calculate the high redshift behaviour of the merger rate of a Poissonian distribution of BHs.Throughout the paper we use $G=c=1$.

\section{The model}\label{sec:1}

\subsection{Motivations}\label{sec:motivations}

As explained in the Introduction, there are good reasons for considering a large set of possible merger rate histories. Here we briefly discuss how these can emerge from clustered initial distribution of PBHs. This
serves as a motivation for the particular templates that we present in section \ref{subsec:themodel}, although the results of the subsequent analysis do not rely on this specific mechanism. In other words, the constraints
on the merger rates we report here are independent of their physical origin.\\

In order to find the merger rate of isolated BBHs, it is necessary to determine the probability $Q(x,y)$ of, given the presence of a BH at $ r=0$, find the two nearest neighbours at positions $x$ and $y$ (where $x$ is position of the binary companion, and $y$ is the position of the BH giving torque to the binary) \cite{Nakamura:1997sm}. This probability in principle depends on  all the N-point correlation functions determining the probability of finding N-BHs. The reason is that if we want to find only two BHs (or any number) within a certain volumen, we have to consider the probability of not finding N-BHs in that volume. This hardens the task of determining in generality this probability. There are however at least two simple cases where it can be easily determined. These are the Poissonian and the so-called ``separable" case.
	
In a Poissonian distribution,  the probability of finding any BH is independent of the presence of other BHs. Then the probability of finding N-BHs at distances $r_1 ... r_N$ from a BH at r=0, $P_N(r_1,..,r_{N-1})$, is simply N-times the probability of finding a single BH, $ P_1^N$. 
Another simple situation, the separable case, is in which the probability $P_N$ is a simple function of the probability of finding 2 BHs, $P_2$. In particular, if the excess probability $\xi_N$, defined as
\be
\xi_N(r_1,..,r_{N-1})\equiv \frac{P_N(r_1,..,r_{N-1})}{P_1^N}-1 
\ee 
follows the simple law \cite{1986ApJ...305L...5J,Ballesteros:2018swv},	
\be\label{eq:separable_clus}
1+\xi_N(r_1,..,r_{N-1})=\prod_{i=1}^{N-1}(1+\xi_2(r_i))  \ , 
\ee
then the probability $Q(x,y)$ takes the following form
\be\label{eq:mergingrate_N}
Q(x,y)=16\pi^2x^2y^2n(x)n(y)\exp\left[-4\pi \int_{R_\text{BH}}^y n(z) z^2\dd z\right]\Theta(y-x) \ .
\ee
where $n$ is the local comoving number density of BHs, given by
\be
n(r)=\bar{n} (1+\xi(r)) \,
\ee
where now $\xi(r)\equiv \xi_2(r)$, and $\bar{n}$ is the mean comoving number density of BHs.  Note that the above expressions also include the case of a Poissonian distribution, for which $\xi(r)=0$.
%
%The expression \ref{eq:mergingrate_N} has been used in most studies concerning clustered distribution \cite{}, although it should be clear that it is only valid in the regime where the condition \ref{eq:separable_clus} holds \footnote{In some cases where the condition \ref{eq:mergingrate_N} is not accomplished, this expression becomes either a lower or upper bound on the true merger rate \cite{Atal:2020igj}.}.
%
Now, given a certain configuration $(x,y)$, a merger happens at a definite time $\tobs=t(x,y)$. This relation can be inverted to find $y(x,\tobs)$. The merger rate at time $\tobs$ can then be obtained by integrating the probability $Q(x,y(x,\tobs))$ over the interval $(\xm,\xM)$, where  $(\xm,\xM)$ are the minimum and maximum $x$ that might contribute to a merger at time $\tobs$ (their expressions can be found in Appendix \ref{sec:appendix}). Then the merger rate at time $t$ is 
\be\label{eq:mergerrate_2}
\frac{\dd R}{\dd t}=\frac{\bar{n}}{2}\int_{x_{\rm min}}^{x_{\rm max}} Q(x,y(x,t))\Big|\frac{\dd y}{\dd t}(t,x)\Big|dx \ ,
\ee
where the factor $1/2$ avoids overcounting the binaries.

The merger rate depends on the correlation function $\xi(r)$ through $Q(x,y)$. The former is ultimately determined by the physical mechanism dictating the distribution of the BHs. In order to illustrate its effect on the merger rate, here we will test a simple form, in which the correlation is nearly constant up to a scale $k_L$ and decays to zero for larger scales. In particular, we choose the following form 

\be\label{eq:correlation_func}
\xi(r,\xi_0,\alpha, k_L)=\xi_0 \dfrac{e^{\Theta(r,k_L,\alpha)}-1}{e-1} \quad \text{with} \quad \Theta(r,k_L,\alpha) = \dfrac{1}{1+e^{2 \alpha \log(k_L^{-1}r)}} \eqdot
\ee

\begin{figure}[t!]
\centering
\includegraphics[width=0.49\textwidth]{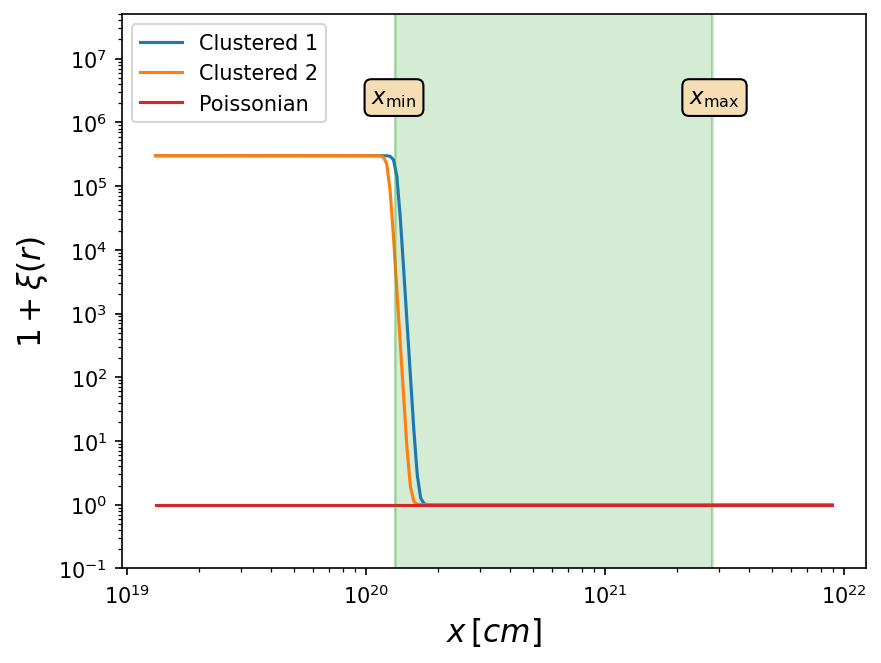}
\includegraphics[width=0.49\textwidth]{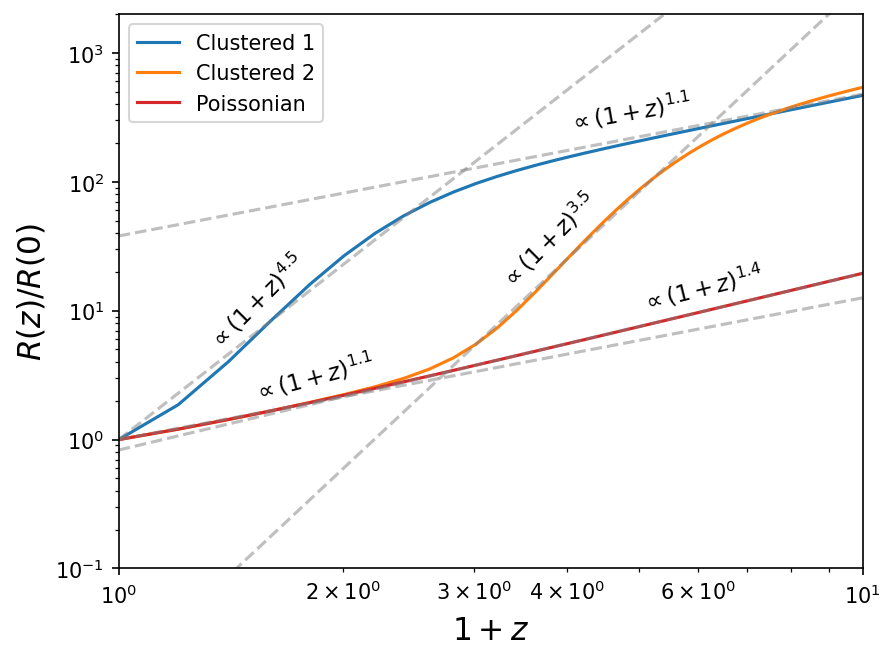}
\caption{left) Three different two-point correlation functions, as a function of comoving distance $x$, resulting in different merger rate histories. In blue and orange we show correlation functions given by Eq. \eqref{eq:correlation_func}, with parameters $(\xi_0,\alpha, k_L)$ given by $(3\times10^{5},30,	0.00887\, k_s)$  and $(3\times10^5,30,0.00959\, k_s)$  respectively. In red, we show the correlation function for a Poissonian distribution, where $\xi(r)=0$. Here $1/k_s$ is the size of 30 solar mass BHs. The comoving scales $x_{\rm min}$ and $x_{\rm max}$ are evaluated at $t=t_{0}$. right) Normalized merger rates for the correlations shown in the left pannel.}
\centering
\label{fig:merger_rates}
\end{figure}

% \be\label{eq:correlation_func}
%\ln \xi(r,\xi_0,\alpha, k_L)=\ln(\xi_0) \left[  \frac{1}{1+e^{-2 \alpha \ln (1 / k_L x)}} -1 \right]  \ .
%\ee
%
The correlation $\xi(r)$ is nearly constant with amplitude $\xi_0$ up to the scale $L=k_L^{-1}$, and then decays to zero with a slope determined by ${\alpha}$. In Fig. \ref{fig:merger_rates}  we show this correlation function (left pannel) and its effect on the merger rate (right pannel). We also show the case of a Poissonian distribution, for which $\xi(r)=0$. As can be seen in Fig. \ref{fig:merger_rates} there can be important changes in the slope of the merger rate, in particular if the correlation function varies around the scales $\xm$ and $\xM$ that determine the scales of the binaries merging at low redshifts. In the examples shown, these changes happen around redshifts 2 to 4.
%s
Note that for the Poissonian case there is a small change in the slope at $z_{*}\simeq 1$, from $1.1$ to $1.4$. While the slope is actually constant in terms of cosmic time $t$, the slope expressed in redshift has a small kink coming from the different scaling of $t$ with respect to $z$.

\subsection{The model}\label{subsec:themodel}

We have presented some motivations for considering changes in the slope of the merger rate of BBH. In the following we will concentrate on its observable consequences. We consider a merger rate given by a broken power law model of the form
\be\label{eq:model}
R(z)=
\begin{cases}
 R_0(1+ z)^\alpha & \text{for } z \leq z_*\\
{\mathcal C} (1+z)^{\beta} & \text{for }  z_* \leq z \leq z_{\rm max} \\
0  & \text{for } z \geq z_{\rm max}
\end{cases}\eqcom
\ee
where $\cal{C}$ is a constant chosen so that the merger rate is a continuous function. The main novelty of the above template with respect to previous works lies in the fact that we consider the posibility of having positive values for both $\alpha$ and $\beta$, which, as shown in the previous section, is indeed possible from the perspective of a population of PBHs. We also note that the above template is a ``sharp" version of the template used in previous works (e.g \cite{Callister:2020arv}). Analysing the effect of a sharp transition in the merger rate is preferable from the perspective of determining precisely the maximum redshift that might influence the stochastic background. For this reason we will mostly deal with the above template for the merger rate, altough we will also comment on the smooth model later on.

We will consider the merger rate given by Eq. \eqref{eq:model} as an extrapolation to the events detected by LIGO/Virgo, such that $R_0$ lies in the range of $9-35$ events/($\rm {Gpc}^3~ yr$)  \cite{LIGOScientific:2021psn}. As for the mass function, we will assume a monochromatic distribution of BHs with mass $M=30 M_{\odot}$, although we will also comment on the effect of a broad mass function.

For the computation of the stochastic background, we will choose $z_{\rm max}=z_{\rm eq}$, where $z_{\rm eq}$ is the redshift of matter-radiation equality. We choose this cutoff since we will consider the contribution of mergers that feature, at some point in their evolution, a circularly inspiriling phase. At higher redshift mergers are very eccentric, and their contribution to the stochastic background might be different from the one considered here. In Appendix \ref{sec:appendix} we show that for a Poissonian distribution this distinction is relevant for mergers happening before or after matter-radiation equality. Since this distinction relies on the background dynamics (more precisely, on the time that a binary at a given distance decouples from the Hubble flow), we  expect a similar behaviour is found for a broader class of initial distributions.

Having said this, it is important to note that all observables related to frequencies in the LIGO/Virgo range are at most sensitive to redshifts $z\sim 10$, and so these predictions are robust to any change in the merger rate happening for redshifts larger than that. On the other hand, we will discuss some theoretical constraints on the parameters of the model that do depend on the redshift of the decay, and so these should be treated with more caution\footnote{For example, in the cases discussed in section \ref{sec:motivations} there might be additional changes in the slope of the merger rate at redshifts $z\gtrsim 10$.}.

Fixing $R_0$ and $z_{\rm max}$, the model is then characterized by three parameters $(\alpha,\beta, z_*)$. Further constraints and priors on the parameters will be discussed in sections \ref{subsec:abundance} and \ref{subsec:priors}.

\section{The stochastic background of past mergers}\label{sec:2}
The energy released by binaries that have already merged contribute to a stochastic background of gravitational waves. Given a merger rate $R(z)$, the energy density of the stochastic background $\Omega_{\rm GW}$ can be expressed in terms of the critical density $\rho_c$ as (see e.g. \cite{Wang:2016ana})
\be\label{eq:omega}
\Omega_{\rm GW} (\nu) =\frac{\nu}{\rho_c H_0}\int_{0}^{z_{\rm sup}}\frac{R(z)}{(1+z)E(z)}\frac{d E_{\rm GW}}{d \nu_s}(\nu_s)dz\eqcom
\ee
where $ d E_{\rm GW}/d \nu_s$ is the GW energy spectrum of the merger and $\nu_s$ is the frequency in the source frame, related to the observed frequency as $\nu_s=(1+z)\nu$. The function $E(z)\equiv H(z)/H_0=[\Omega_r(1 +z)^4+ \Omega_m(1 +z)^3+ \Omega_\Lambda]^{1/2}$. The energy released in GWs considering inspiral, merging and ringdown phases is given by  \cite{Ajith:2007kx,Ajith:2009bn}
\be\label{eq:dEdnu}
  \frac{d E_{\rm GW}}{d\nu_s}(\nu_s) = \frac{\pi^{2/3}M_c^{5/3}}{3}%\left.
  \begin{cases}
    \nu_s^{-1/3} & \text{for } \nu_s < \nu_1 \\
    \omega_1\nu_s^{2/3} & \text{for } \nu_1 \leq \nu_s < \nu_2 \\
    \omega_2\frac{\sigma^4\nu_s^2}{\left(\sigma^2+4\left(\nu_s-\nu_2\right)^2\right)^2} & \text{for } \nu_2 \leq \nu_s < \nu_3 \\
    0 & \text{for } \nu_3 \leq \nu_s
    \end{cases}\eqcom
  %\right\} 
\ee
where $\nu_i\equiv(\nu_1,\nu_2,\sigma,\nu_3)=(a_i \eta^2+b_i \eta + c_i)/(\pi M)$,  $M=m_1+m_2$ is the total mass, $M_c$ is the chirp  mass ($M_c^{5/3}=m_1m_2M^{-1/3}$),  and $\eta=m_1m_2 M^{-2}$ is the symmetric mass ratio. The parameters $a_i$, $b_i$ and $c_i$ can be found in \cite{Ajith:2007kx}, and ($\omega_1, \omega_2$) are chosen such that the spectrum is continuous. For BHs of $30\, \Ms$ the characteristic frequencies are $\nu_i=(135,271,79,387) \text{ Hz}$.  The upper limit of the integral in Eq.\eqref{eq:omega} is given by $z_{\rm{sup}} =\text{min}(z_{\rm{max}}, \nu_{3}/\nu-1)$, since for $z\geq z_{\rm{sup}}$, either the merger rate or the energy density are zero. Note that there might also be additional contributions to the stochastic background coming from (not bounded) hyperbolic encounters \cite{Garcia-Bellido:2021jlq}.\\

\begin{figure}[t!]
\centering
\includegraphics[width=0.49\textwidth]{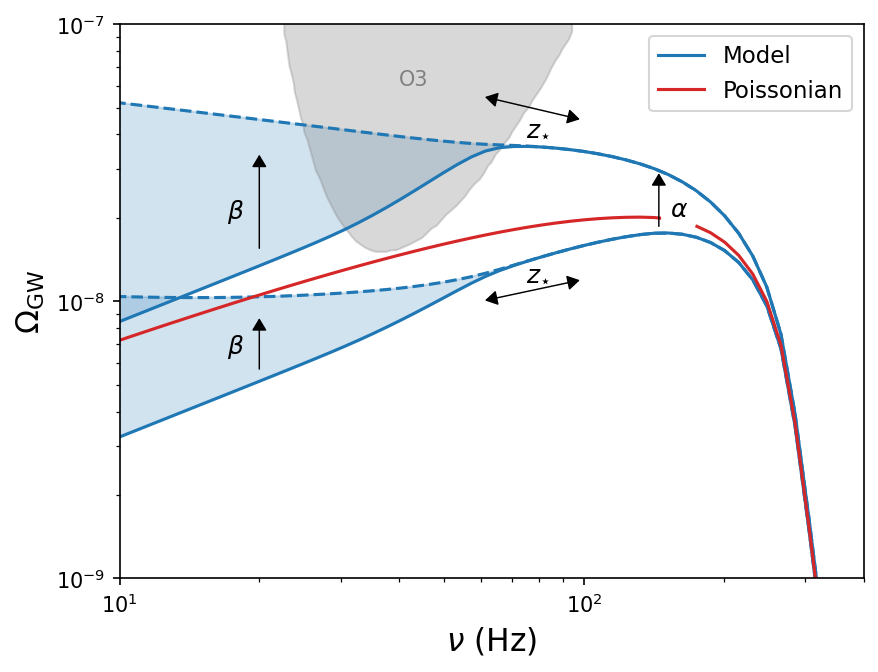}
\includegraphics[width=0.49\textwidth]{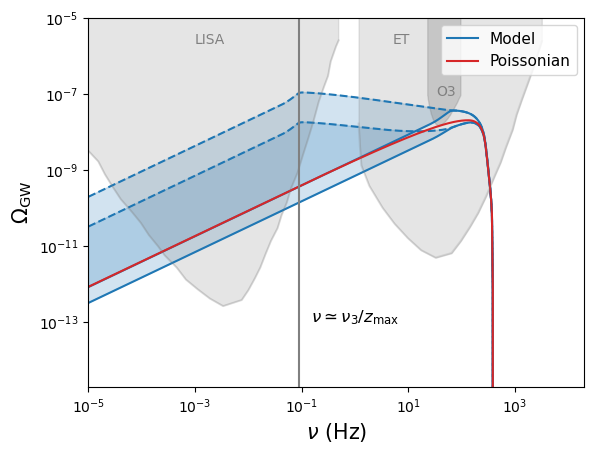}
\caption{Left) Gravitational wave background for the merger rate as given by the broken power law model, for BHs of $30 \Ms$, and varying parameters ($\alpha, \beta, z_*$). Here we show the LIGO/Virgo range of frequencies. The parameters for these curves are $R_0=9$, $\alpha\in(0.8,3)$, $\beta\in(-5,2.7)$, and $z_*=3.5$. The arrows indicates the effect of varying these parameters. The red curve has $\alpha=1.1$, $\beta=1.4$ and $z_*=1$, and describes the case of an initial Poisson distribution of BHs with no binary disruption.
Right) Gravitational wave background with the same parameters, expanded to include the LISA frequency range. For the values of $z_*$ under consideration, the signal in the LISA region follows the $2/3$ scaling with an amplitude depending on ($\alpha,\beta,z_*$). Let us note that at LISA scales there might be additional contributions to the stochastic background coming from extremely eccentric orbits (where the merger time is given by the free-fall time) that we do not consider here.}
\centering
\label{fig:omegas}
\end{figure}

In Fig. \ref{fig:omegas} we show the corresponding gravitational wave signal for a merger rate given by Eq. \eqref{eq:model} for a range of the parameters ($\alpha,\beta,z_*$). We also show the behaviour of $\Omega_{GW}$ resulting from a Poissonian initial distribution of PBHs.  In this case we identify the well known $\nu^{2/3}$ scaling at small frequencies with the posterior decay at larger frequencies. Once we allow for variations in the merger rate, the family of possible spectra enlarges.

Generally, $\alpha$ and $\beta$ set the overall amplitude at large and small frequencies respectively, and the frequency of the transition, $\nu_*$, is controlled by $z_*$ (more precisely $\nu_{*}=\nu_{3}/(1+z_*)$). For $\nu \geq \nu_{*}$, the only contribution to $\Omega_{\rm GW}$ comes from redshifts up to $z_*$. Therefore, the exponent $\alpha$ alone dictates the behaviour of $\Omega_{\rm GW}$ for this frequency range. On the other hand, for $\nu \leq \nu_{*}$, all redshifts $z\leq z_{\rm{sup}}$ contribute. For very small frequencies the main contribution comes from redshifts $z\geq z_*$, so it is the exponent $\beta$ that governs the behaviour of $\Omega_{\rm GW}$. In principle, the infrared (IR) tail of the signal follows the scaling
\be\label{eq:law}
\Omega^{\rm IR}_{\rm GW}(\nu,z_{\rm max}\rightarrow \infty)\sim \begin{cases}
  \nu^{2/3}\eqcom \beta <\frac{7}{3}\\
  \nu^{3-\beta}\eqcom \beta \geq \frac{7}{3}
\end{cases}
\ee
where the value $\beta_{\rm thr} = \frac{7}{3}$ comes from the fact that the integral in Eq.\eqref{eq:omega} does not converge at large redshifts for $\beta>\beta_{\rm thr}$. In practice, we found that due to the cutoff of the merger rate at $z_{\rm max}=z_{\rm eq}$\footnote{See the discussion of this point in the Appendix \ref{sec:appendix}.}, this scaling is not achieved in our model (would the cutoff be at larger redshift, then the scaling would be recovered). Since the signal is however tending to this limit, there is an intermediate region not necessarily given by the well known $2/3$ scaling of the IR. Moreover, even when the 2/3 scaling is achieved, the amplitude of the signal in this region is dependent on $\beta$. The strong dependence of the SGWB on the parameters describing the merger rate shows that these might be strongly constrained. In order to determine this precisely, we compute the signal-to-noise ratio of the SGWB as a function of the parameters $\alpha,\beta,z_*$ and for a number of experiments.

\section{The Signal-to-Noise Ratio and the abundance criterion}\label{sec:3}
For a merger rate density $R(z)$, we compute the stochastic background $\Omega_{\rm GW}(\nu)$, and the signal-to-noise ratio (SNR) that it generates in some detector. For detecting a stochastic background we need to rely on two different detectors, otherwise the noise overkills any underlying signal (unless the amplitude is excessively large).  Then, the SNR is given by
\be
\text{SNR}=\sqrt{T_{\rm obs} \int^{\nu_{\rm max}}_{\nu_{\rm min}} d\nu \left(\gamma(\nu)^2\frac{\Omega_{\rm GW}(\nu)}{\Omega_{\rm exp}(\nu)}\right)^2} \ ,
\ee
where $T_{\rm obs}$ is the time of observation,  $\gamma(\nu)$ is the so-called overlap reduction function between the two detectors (that can be found in \cite{Flanagan:1993ix}), and $\Omega_{\rm exp}(\nu)$ is the noise energy density in terms of the critical density, given by \cite{Romano:2016dpx}
\be
\Omega_{\rm exp}(\nu)=2\pi^2\nu^3 S_n(\nu) / (3 H_0^2) \ ,
\ee
where $S_n(\nu)$ is the experimental strain, in units of strain/Hz. It is related to the amplitude spectral density $h^{1,2}_{\text{\rm eff}}(\nu)$  of detectors $1,2$ as
\be
h^{1}_{\text{\rm eff}}(\nu) h^{2}_{\text{\rm eff}}(\nu)=S_n(\nu) \ .
\ee
For the case of LIGO/Virgo, we use a representative of O3 for the amplitude spectral density $h^{1,2}_{\text{\rm eff}}(f)$ of Hanford and Livingston detectors that can be found in (https://dcc.ligo.org/LIGO-T1500293/public).  We use the fact that the third run detected no stochastic signal, in a time interval of 205.4 days of coincident observations of the Hanford and Livingston detectors \cite{KAGRA:2021kbb}. For future constraints we use the Advanced LIGO (aLIGO) design configuration, the Einstein Telescope (ET), LISA and SKA. The amplitude spectral density data for these experiments, except LISA and SKA, can be found in (https://dcc.ligo.org/LIGO-T1500293/public). For SKA we use the data from \cite{Moore:2014lga}, while for LISA we follow the procedure explained in \cite{Smith:2019wny}. In this case $\Omega_{\rm exp}(\nu)$ is given by
\be
\Omega_{\rm exp}=\Omega_{\rm I}\frac{4\pi^2\nu^3}{3H_0^2} \quad \rm{with} \quad \Omega_I=\sqrt{2}\frac{20}{3}\left[\frac{S_I(\nu)}{\left(2\pi \nu\right)^4}+ S_{II}\right]\left[1+\left(\frac{\nu}{4\nu^\star/3}\right)^2\right]\eqcom 
\ee
where $\nu^\star=c/2\pi L$ with $L=2.5\times 10^6$ km and the functions $S_I$ and $S_{II}$ are given by
\be
S_I=5.76\times 10^{-48}\left(1+(\bar{\nu}/\nu)^2 \right)~ {\rm Hz}^{3}  \quad \rm{with} \quad  S_{II}=3.6\times 10^{-41} ~{\rm Hz^{-1}}\eqcom
\ee
where $\bar{\nu}=0.4 \text{ mHz}$ and we take the effective time of observations to be $T_{\rm obs}$=3 yr. Let us note that we expect many sources of GWs acting as foregrounds to this signal (for a review on stochastic sources, see \cite{Christensen:2018iqi}), and so a more refined estimation of the SNR should also take into account the ability to separate these components  (see e.g. \cite{Moore:2019pke,Karnesis:2021tsh}).

With these functions at hand, we can proceed to calculate the SNR for the aforementioned experiments. We set the threshold for detection to be $\text{SNR}=2$, and first use the fact that LIGO/Virgo O3 detected no signal to constrain the triplets $(\alpha, \beta, \text{z}_*)$  producing a signal with $\text{SNR}>2$. 
%Since the SNR for future experiments turns out to be quite large (except at PTA scales), we only show the minimum of the SNR, corresponding for our model to the SNR for a merger rate with $\alpha=0.8$,$\beta=-5$, $z_*=0.8$). Before showing these results, we will comment on the space of parameters $(\alpha, \beta, z_*)$ under scrutiny.

\subsection{The abundance constraint}\label{subsec:abundance}

From the merger rate $R(z)$ we can find the total number of black holes in the form of binaries. If we define $f_{\rm b}$ as the fraction of black holes in binaries and as $f_{\rm BH}$ the total amount of dark matter in the form of black holes, then the number density of black holes in binaries is given by
\be\label{eq:_npbh}
\int_{0}^{z_{\rm max}} \frac{R(z)}{H_{0}(1+z)E(z)} dz = f_{\rm b} f_{\rm BH} \frac{n_{\rm DM}}{2}. 
\ee
We expect the fraction of BH in binaries to be quite small. In some simulations of BH clusters it has been shown that $f_{\rm b}\simeq \mathcal{O}(10^{-3})$ \cite{Trashorras:2020mwn}. If we adopt this as a fiducial value, and because $f_{\rm BH}\leq1$ then it holds that
\be\label{eq:constraint_npbh}
\int_{0}^{z_{\rm max}} \frac{R(z)}{H_{0}(1+z)E(z)} dz \leq 0.001\, n_{\rm DM} \ .
\ee
This equation will allow us to constrain the possible values of $({\alpha,\beta,z_*})$. In principle there might be some additional observational constraints on the total number of primordial black holes. In this range of masses, these come from the non-observation of microlensing events in EROS \cite{EROS-2:2006ryy}, MACHO \cite{Macho:2000nvd}, and type Ia supernovae \cite{Zumalacarregui:2017qqd}. These constraints can however be evaded if the BHs sourcing the lenses are clustered \cite{Garcia-Bellido:2017xvr}, and so we do not consider them here.%

\subsection{Priors for the model parameters}\label{subsec:priors}

As for the parameter space, we choose values for the transition redshift $z_* \geq 0.8$. That is, we consider that the change in the slope of the merger rate happens for redshifts larger than the redshift of the most distant binary merger that has been directly observed. This ensures that the current constraints on $\alpha$ coming from the direct observations also hold in this model. In particular, this means that $\alpha=2.7^{+1.8}_{-1.9}$ \cite{LIGOScientific:2021psn}, and that we can treat the constraints coming from direct and indirect detection as being independent. We will then choose $\alpha$ in the interval $(0,5)$, keeping in mind that values $\alpha<0.8$ are outside the $90\%$ confidence level found from the analysis of direct observations  {\cite{LIGOScientific:2021psn}.

As we already mentioned we will allow $\beta>0$, since we are interested in exploring new merger histories for the case of PBHs.  The case with $\beta<0$ describes astrophysical BHs \cite{Madau:2014bja} (for an analysis of the stochastic background which such condition, see \cite{Callister:2020arv}, and for the estimation of the evolution of the merger rate using direct observations, see \cite{Fishbach:2018edt}). We will then choose $-5<\beta<5$. Let us note that for positive values of $\beta$, the abundance criteria Eq.\eqref{eq:constraint_npbh} result in a restriction $\beta \lesssim 2.3$, roughly independent of $\alpha$ and $z_*$ (for $z_*\ll z_{\rm max}$). For smaller allowed fractions of binaries with respect to DM (e.g. $f_{\rm b}<0.001$), the bound on $\beta$ changes slightly (by $10\%$ with respect to  $f_{\rm b}<1$).
We will choose $R_0$ as the minimum allowed by direct detection, i.e $R_0=9$, and so the constraints presented here are a lower bound on the SNR within this model.  Since we choose $R_{0}$ consistent with the current present merger rate, we implicitly assume that the sum of all BHs (astrophysical and primordial) is given by such power law. In this sense, the change in the slope at $z_*$ could also represent a change in the binary population, e.g. a transition from astrophysical to primordial binaries. Of course, following the same type of reasoning, we should also take into account the change in the spectrum of masses resulting from each channel of BH formation. It is not our goal here to perform such study, but to delve into the importance of the stochastic background to determine changes in the merger rate, and so we will consider the simple scenario of a single mass function.  

\section{Results}\label{sec:4}

In the following we proceed to present contours of the SNR as a function of the parameters $\alpha$, $\beta$ and $z_*$.  Note that since the shape of the signal (and not just the overall amplitude or scale) changes with the varying parameters, it is probably not very accurate to compare our model with a single integrated sensitivity curve (which is constructed assuming a signal with a single power law), as can be done e.g. in \cite{Thrane:2013oya,Schmitz:2020syl}. For the baseline cosmological model we use the results of the Planck satellite \cite{Planck:2018vyg}.

\subsection{LIGO/Virgo - O3 run  - Present Constraints}

In Fig. \ref{fig:ligo} we show the SNR for the LIGO/Virgo O3 run. The region in grey corresponds to sections of the parameter space where SNR$>2$, and thus can be considered ruled out by the non-observation of a stochastic background. The green dashed line shows the equality of the abundance constraint Eq. \eqref{eq:constraint_npbh}, so all the parameters above that line are also ruled out. 

From these results we can see that LIGO/Virgo O3 is sensitive to the changes in the stochastic background coming from the evolution of the merger rate at large redshift ($4>z>0.8$). On the other hand, the SGWB is not very sensitive to changes in the merger rate happening at redshift $z_*>4$. This shows that using the stochastic background we can constrain the merger rate at redshifts larger than the horizon redshift for the detection of individual mergers, $z_{\rm hor}\simeq1$. 

In general, we find that the non detection of a stochastic background largely restricts the allowed parameter space. We find that for any $\beta$ and for $z_*>0.8$, $\alpha<3.7$. The maximum allowed value for $\alpha$ decreases as the redshift of transition increases. For $z_*>4$, we found $\alpha<1.3$, which combined with the direct observation constraint implies that $\alpha$ lies in the small range $0.8<\alpha<1.3$. As for $\beta$, we found that it is also greatly constrained. For transitions happening at redshift $z_{\ast}<1.6$, we find $\beta<2.3$ from the non-detection of a SGWB. For transitions happening at larger redshifts, larger values of $\beta$ are allowed from the perspective of the SGWB, but then either the abundance constraint or the direct detection constraint on $\alpha$ is not satisfied. 
\begin{figure}[!ht]
\includegraphics[width=0.49\textwidth]{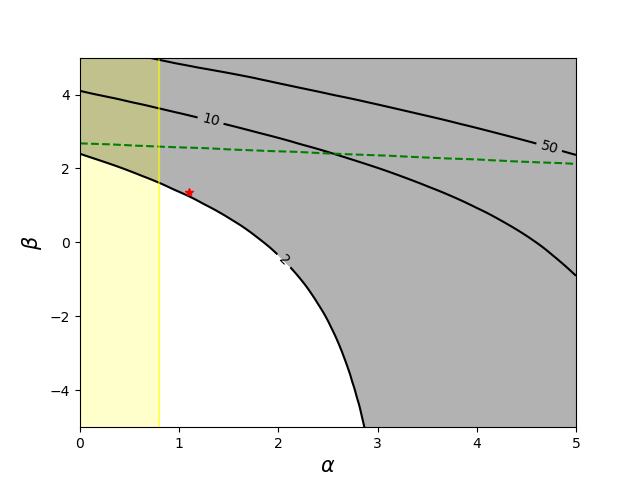}
\includegraphics[width=0.49\textwidth]{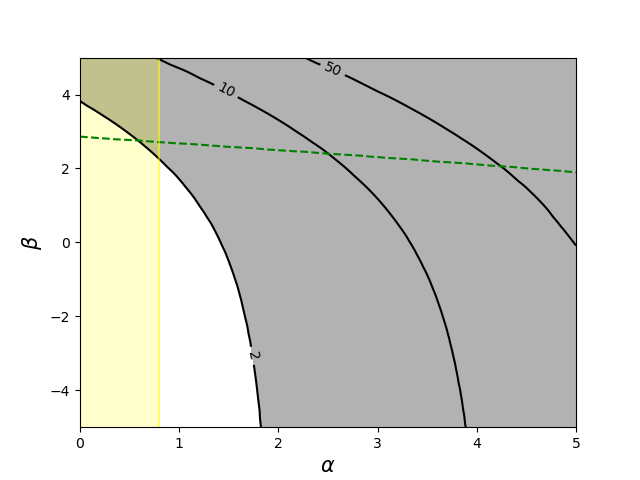}
\includegraphics[width=0.49\textwidth]{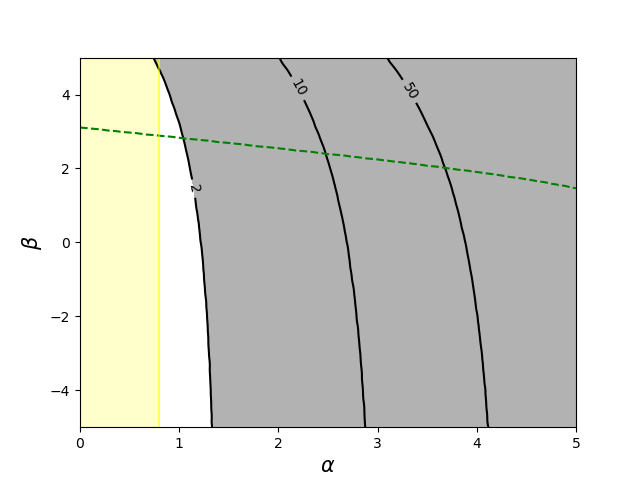}
\caption{SNR for LIGO/Virgo O3 run and for our model. At the top left $z_* =1$ (where we highlight in red the Poissonian case, $\alpha=1.1$ and $\beta=1.4$), while at the top right $z_* = 2$ and at the bottom $z_* = 4$. The SNR becomes independent of $\beta$ when the transition $z_*$ happens for $z>4$. For $z_*<4$, the constraints on the high redshift evolution ($\beta$) are similar to the low redshift evolution ($\alpha$). The green line is the equality of the abundance criterion, given in Eq. \eqref{eq:constraint_npbh} and with $z_{\rm max}=z_{\rm eq}$ (practicaly the same bound is found for $z_{\rm max}>z_{\rm eq}$) and the yellow band is the exclusion of $\alpha$ coming from direct measurements.
}
\label{fig:ligo}
\end{figure}

\subsection{The Poissonian PBH case}
An interesting case is when the merger rate is given by a population of PBHs following an initial Poissonian distribution. The simplest way to constrain this scenario is with the total number of events. For $R_0=9-35 ~\rm{Gpc^{-3}yr^{-1}}$, this implies that $f_{\rm PBH}<0.001-0.003$ (considering $R_0$ as the maximum possible number of PBH binary mergers)\footnote{In this case, an analytical expression for the merger rate can also be obtained \cite{Raidal:2017mfl,Sasaki:2018dmp}.}.  Interestingly however, the stochastic background provides a competitive constraint on the abundance (see \cite{Wang:2016ana} for a previous analysis along these lines). In this case, $\alpha=1.1$, $\beta=1.4$ and $z_*=1$ (shown in red in Fig. \ref{fig:omegas} for $R_0=9 ~\rm{Gpc^{-3}yr^{-1}}$). For these parameters SNR $<$ 2 implies $R_0<8$, which from the expression above implies $  f_{\rm PBH} < 0.001$. This shows that the constraints coming from the stochastic background are  of the same order, or even stronger, than the constraint coming from direct observations.

\subsection{Future constraints in LIGO/Virgo and LISA band.}

While the O3 constraints still leave some part of the parameter space viable,  future runs of LIGO/Virgo/Kagra and future GW experiments operating at frequencies larger that $10^{-4}$ Hz should be able to detect a signal for this family of merger rates. In Fig. \ref{fig:ET_stochastic} (left pannel,  black curve) we show the smallest SGWB background from this model that is allowed by O3 (i.e $\alpha=0.8,\beta=-5, R_0=9, z_*=1$)\footnote{For small $\beta$, the SGWB becomes independent of $\beta$. In this sense $\beta=-5$ is very similar to considering a sharp cutoff ($\beta\rightarrow -\infty$).}.   

For any allowed parameter, the signal should thus be visible  by future runs of LIGO/Virgo, and also by LISA, which operates at smaller frequencies than LIGO/Virgo, around $(10^{-4},10^{-1})$ Hz. This interplay between observations in LIGO/Virgo and LISA bands has already been discussed for PBHs \cite{Clesse:2016ajp,Chen:2018rzo,Wang:2019kzb,Chen:2021nxo}. 

Even though at intermediate frequencies, $\nu \in (0.1,10)$ Hz, the slope of the SGWB shows large deviations with respect to the canonical $2/3$ scaling,  in the LISA frequency range the $2/3$ scaling is recovered. The amplitude at these scales however depends on the parameters ($\alpha,\beta$). 

These considerations imply that if we only consider observations in the LISA band there might be a degeneracy between ($\alpha,\beta$) and the mass of the binary system, since BBHS with larger masses results in spectra radiating at smaller frequencies. This degeneracy is however broken by comparing the signal at both LIGO/Virgo and LISA bands.

Naturally, experiments planned at frequencies between LISA and LIGO/Virgo, as Einstein Telescope \cite{Punturo:2010zz}, DECIGO \cite{Kawamura:2006up} and TianQin \cite{TianQin:2015yph}  will also contribute in a similar direction. These experiments have the advantage of being sensitive in the range of frequencies where the signal deviates from the $2/3$ scaling.

\subsection{Pulsar Timing Arrays}

The infrared tail of the SGWB might be largely enhanced with respect to the Poissonian case and thus there might be consequences for observables at very small frequencies. In particular, it is interesting to compute the predictions at pulsar timing array scales, lying in a frequency interval $\nu \in (10^{-9}-10^{-6})$ Hz.  

The signal detected by the NANOGrav collaboration \cite{NANOGrav:2020bcs} has already been shown to be accommodated by the IR tail of the stochastic background coming from mergers of super (or "stupendously") massive BHs, either of astrophysical \cite{Middleton:2020asl} or primordial origin \cite{Atal:2020yic}.  We found however that for BBHS of $30 M_{\odot}$  the enhancement is not so pronounced to neither explain the NANOGrav signal nor being detectable by SKA (and at the same time being consistent with LIGO current bounds and the abundance constraint) \footnote{Note that since we are extrapolating to very small frequencies, the first thing is to determine the smallest possible frequency generated by a merger.  In particular, for a spectrum of the form given by Eq.\eqref{eq:dEdnu}, it is assumed that the frequency spectrum has a tail $\nu^{-1/3}$ in the IR, without any additional cutoff. We should however consider that at any time there is a minimum frequency at which a binary can radiate, given by the maximum distance at which they could be separated. This in turn is given by the radius at which they decouple from the Hubble flow. This is found to be $\nu_{\rm min}(z) =  (1+z)^{1/3}\left(\frac{M}{\rho_{\rm eq}a_{\rm eq}^4}\right)^{-1/3}$, which is smaller than PTA scales for the masses considered.}. It might be possible however that for BBH of larger masses, the signal at these scales is large enough to be detectable. 
%
%The first thing is to determine the smallest possible frequency generated by a merger. Note that in \ref{eq:dEdnu} we assume that the frequency spectrum has a tail $\nu^{-1/3}$ in the IR, without any additional cutoff. We should however consider that at any time there is a minimum frequency at which a binary can radiate, given by the maximum distance at which the*y could be separated. This in turn is given by the radius at which they decouple from the Hubble flow. The radius $R_{\rm max}$ at which a binary decouples is found from the condition that the energy density of the a binay with major semiaxis $R_{\rm max}$ is larger than the background energy density, i.e.
%\be\label{eq:dec}
%\frac{M}{R_{\rm max}(t)^3}=\rho_{\rm back}(t) \ .
%\ee
%Note that after matter radiation equality, both sides of \eqref{eq:dec} have the same scaling with time, meaning that any binary should decouple before matter radiation equality. 
%At these early times, we find
%\be
%R(t)=\left(\frac{M}{\rho_{\rm eq}}\frac{a(t)^4}{a_{\rm eq}^4}\right)^{1/3} \ .
%\ee
%The smallest possible observed frequency is then 
%\be
%\nu_{\rm min}(z) = c (1+z)^{1/3}\left(\frac{M}{\rho_{\rm eq}a_{\rm eq}^4}\right)^{-1/3} \ .
%\ee
%The minimum observed frequeny scales as $(1+z)^{1/3}$, implying that the largest minimum frequency comes from most early mergers. These corresponds to mergers happening at the time of PBH formation, and for these $\nu_{\rm min}= 10^{-9}$ Hz. This implies that we can adopt the template \eqref{eq:dEdnu} for such low frequencies.\\

%Now we can compare with present and future PTA experiments.

\subsection{A joint analysis with direct detection: Einstein Telescope}\label{sec:ET}

Here we discuss in more detail the prospects for making a joint observation of the stochastic background and individual sources. For mergers of this mass, the Einstein Telescope (ET) should detect individual mergers, up to redshifts $z\simeq 10$ \cite{Maggiore:2019uih}.

In order to better assess its constraining power, lets take two sets of parameters having the same SNR in the stochastic background in the LIGO experiment, for example ($\alpha,\beta$)=(0.7,0.2) and ($\alpha,\beta$)=(0.2,0.7), with $z_*=0.8$ in both cases. These two sets of parameters generate an almost identical stochastic background, as can be seen in the Fig. \ref{fig:ET_stochastic}. However they could be distinguishable if  the merger rate as a function of redshift can be inferred from direct observations. We can estimate the number of mergers observed by the ET by using the expression
%Using the results from Fig.  (2) of \cite{Maggiore:2019uih} we  estimate the probability of detecting these type of mergers in the Einstein Telescope by the simple expression 
%\beq
%P_{ET}(z) = \begin{cases}
%1  & \quad z \le z_{\text{ET}} \\
%0 & \quad  z > z_{\text{ET}}
%\end{cases}
%\eeq
%where we will take $z_{\text{ET}}\approx 10$ (these are just rough estimates but should suffice for our purposes). The total number of mergers detected by ET will be given by,
\beq
N_{\text{obs}} (z) = \int_{0}^{z} {dz' ~R(z') \left(\frac{d V_c}{dz'}\right) \left(\frac{T_{\text{obs}}}{1+z'}\right) P_{ET}(z')}~,
\eeq
where we have denoted by $V_c$ the comoving volume given by,
\beq
\left(\frac{d V_c}{dz}\right)  = 4\pi \left(\frac{1}{H_0}\right)^3 \left(\int_0^z{\frac{dz'}{E(z')}}\right)^2 E(z)^{-1}~,
\eeq
%where
%\beq
%E(z) \approx \left( \Omega_M (1+z)^3 + \Omega_{\Lambda}\right)^{1/2}
%\eeq
and we take the probability of observation of these mergers by the ET to be $P_{ET}(z)\approx 1$ for $z<10$. Finally, we take the observation time, $T_{\text{obs}}$, to be $1$ year.
Using the expressions above we can find the number of expected mergers as a function of redshift. In Fig. \ref{fig:ET_stochastic} we show the number of sources that ET should detect, showing that both models generate large differences in the number count. This means that we can use this  measurement to break the degeneracy between models whose signal to noise ratio for the stochastic background is the same. 
\begin{figure}[!t]
\includegraphics[width=0.47\textwidth]{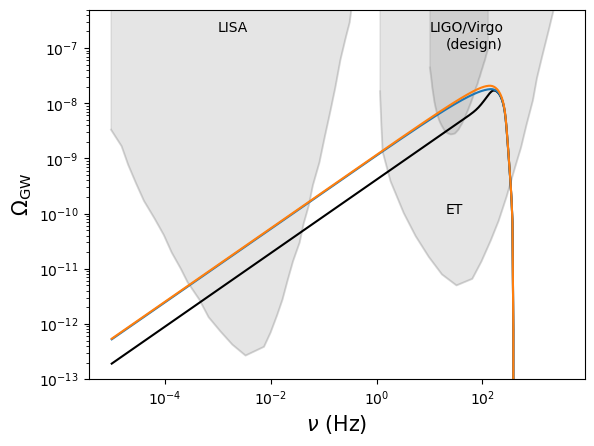}
\includegraphics[width=0.485\textwidth]{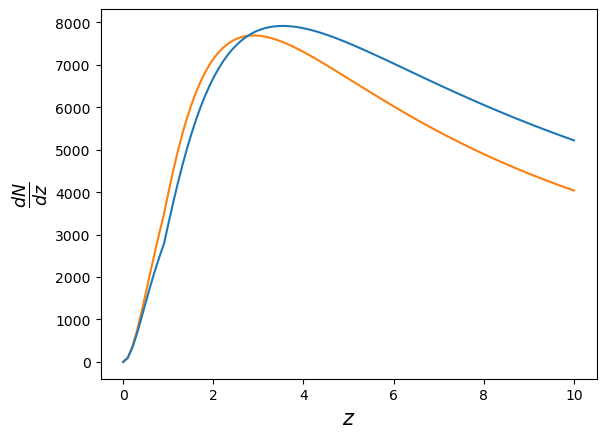}
\caption{left) SGWB for two different models generating a very similar stochastic background and characterized by the
following parameters. Model 1 (blue): $\alpha=0.5$, $\beta=0.25$ and $z_{\text{max}}=0.8$ and Model 2 (orange): $\alpha=0.25$, $\beta=0.5$ and $z_{\text{max}}=0.8$. In black we show the smallest stochastic background generated by this class of models ($R_0=9$, $\alpha=0.8$, $\beta=-5$ and $z_*=0.8$). This means that the signal should be observed among other observatories, by LIGO/Virgo design configuration, ET and LISA. right) Number of expected mergers, for the same value of the parameters. While the stochastic background are nearly degenerated, there are large differences in the expected number of events in the ET.}
\label{fig:ET_stochastic}
\end{figure}

\subsection{Broad mass function and smooth transition models}

In this section we will briefly comment on the effect of broad mass functions and a smooth merger rate transition on the detectability of a given model.
As for the mass function we will consider a log-normal distribution \cite{PhysRevD.47.4244,Kannike:2017bxn}, given by 
\be
P(m;M_c,\sigma)=\frac{1}{m\sigma\sqrt{2\pi}}e^{-\left(\ln(M/M_c)^2/2\sigma \right)} \ .
\ee 
The stochastic background coming from mergers following such distribution can be calculated from the mean energy emitted by a pair of such BHs. This is obtained by taking many pairs of BHs from the above distribution, calculate the energy released by all such pairs, and average. We show, in Fig. \ref{fig:Enu}, the mean energy emitted  by a distribution in comparison with the energy emitted by a pair of BHs having the same mass. The effect of considering such distribution is to broaden the peak towards larger frequencies, while the amplitude slightly decreases. This, in turn, creates a similar effect at the level of the stochastic background. The resulting background is calculated with the same expression (\ref{eq:omega}) but replacing $dE_{\rm GW }/d \nu_s$ with the mean $\langle dE_{\rm GW}/d \nu_s \rangle$. For the masses under consideration, the effect of the distribution is to slightly lower the amplitude of the signal at LIGO frequencies. The opposite would be true for more massive binaries.
\\
\\

\begin{figure}[!t]
\includegraphics[width=0.49\textwidth]{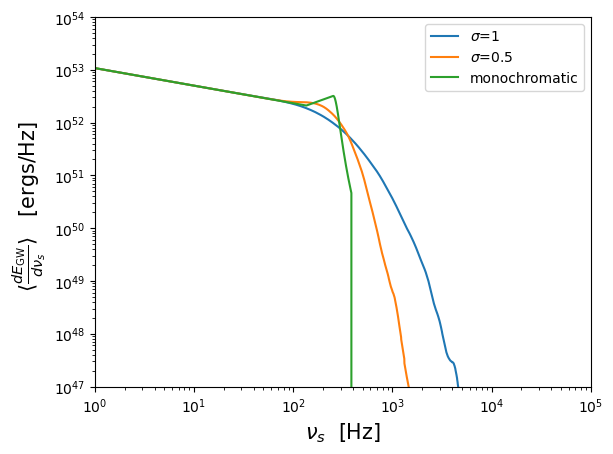}
\includegraphics[width=0.49\textwidth]{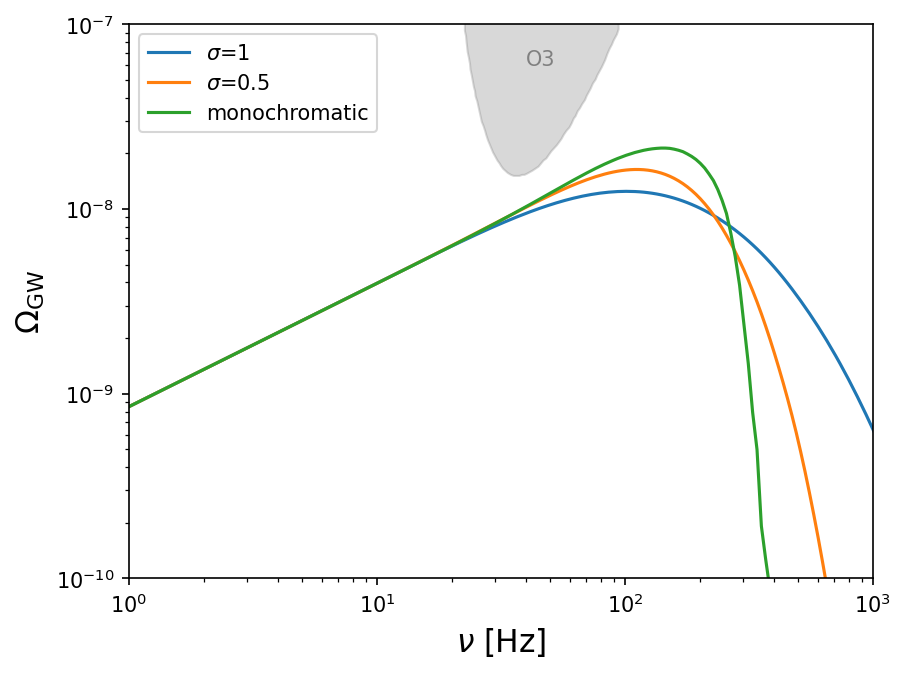}
\caption{left) Comparison of the mean energy emitted by a binary BH between a monochromatic and a broad distribution of masses. The chosen values of $\sigma$, between 0 and 1, are reasonable from the point of view of the mergers that haven been measured \cite{DeLuca:2021wjr}. right) Effect on the stochastic background.}
\label{fig:Enu}
\end{figure}

As for a smooth merger rate model, we show results for the model used in \cite{Callister:2020arv}, given by
\be
R(z)=R_0\frac{(1+z)^{\alpha}}{1+\left(\frac{1+z}{1+z_*}\right)^{\alpha-\beta}} \ .
\ee
This model, similarly to the one we have studied, also behaves as $(1+z)^{\alpha}$ at small redshifts, and as $(1+z)^{\beta}$ for large redshifts. The difference is that the amplitude of the merger rate for the smooth template is larger at small redshifts with respect to the sharp template, while at large redshifts the opposite is true. This in turn means that the constraints are weaker for $\alpha$ and stronger for $\beta$ with respect to the sharp template. We show an example of this in Fig. \ref{fig:SNR_comp}.

\begin{figure}[!t]
\includegraphics[width=0.49\textwidth]{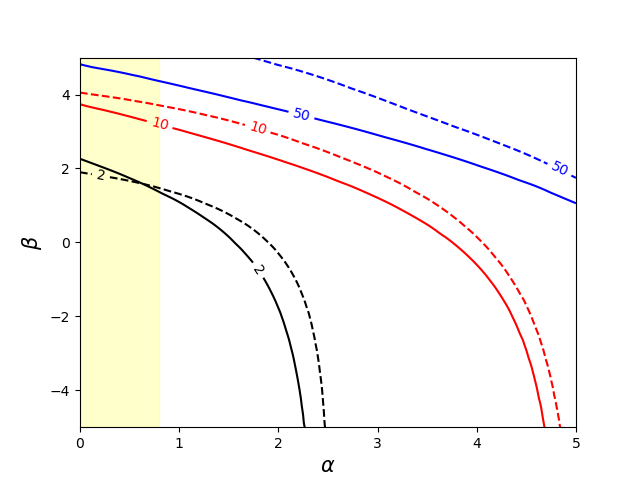}
\includegraphics[width=0.49\textwidth]{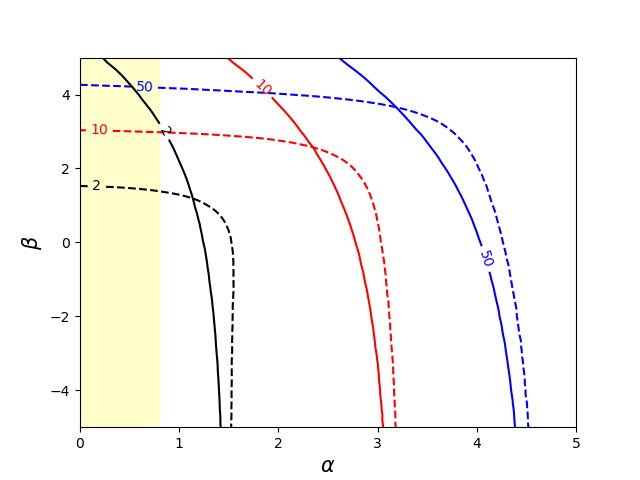}
\caption{left) Comparison of the SNR coming from two templates having the both a behaviour $(1+z)^{\alpha}$ and $(1+z)^{\beta}$ at small and large redshifts. In solid lines we show the model we use in this paper which is given by a sharp cutoff. In dashed lines we show the model most used in the literature, which smoothly interpolates between the two regimes. The two figures correspond to two different values of $z_*$, given by $z_*=0.8$ and $z_*=3.5$ in the left and right plot respectively.}
\label{fig:SNR_comp}
\end{figure}

\section{Discussion and Conclusions}\label{sec:conclusions}

We have tested a model inspired by a population of PBHs, which can have an increasing merger rate up to large redshifts. We have shown that if the merger rate exhibits changes in its slope, a wide range of possible stochastic backgrounds can be generated.

In particular, we have shown that in a mid-frequency range of the SGWB, between $\nu \in (\nu_3/z_{\rm max},\nu_3)$ and where $z_{\rm max}$ is the redshift at which the merger rate decays, the spectra can exhibit slopes that are different from the usually expected $\nu^{2/3}$ scaling\footnote{Note that this is only the case for value of $\beta>0$.}. At smaller frequencies the $\nu^{2/3}$ scaling is recovered, however the amplitude turns out to be highly dependent on the slope of the merger rate. This dependence shows that the SGWB is a powerful probe of the merger history of BBHs.

We have thus analysed to which extent the current non-detection of a SGWB constrains the merger rate. We have found that indeed the non-observation of the SGWB puts constraints on the merger rate at redshift as high as $z\sim 4$, a redshift that is much larger than the maximum redshift at which we have detected BH binaries. We have also shown that future runs of LIGO/Virgo should be able to detect such SGWB and that the signal should be visible by LISA for any of the considered parameters of the model.

We have finally considered the prospects of Einstein Telescope for determining the merger rate through the observations of individual mergers. The expected number of BBHs is so large that  the existing degeneracies in the stochastic background can be broken, meaning that the merger rate can be unambiguously determined.

Finally, we should mention that a large merger rate at high redshifts should also imply that a large number of the events seen by LIGO/Virgo would be lensed. This is a certain possibility \cite{Mukherjee:2020tvr,Diego:2021fyd} although more data is needed in order to confirm such hypothesis. This means that lensing provides a complementary channel for the determination 
of the merger rate.

\section*{Acknowledgments}
We thank Heling Deng and Guillem Dom\`enech for discussions and comments on the manuscript.
This work was supported by the Spanish Ministry MCIU/AEI/FEDER grant (PGC2018-094626-BC21) and the Basque Government grant (IT-979-16).

\appendix

\section{The high redshift merger rate for a Poissonian initial distribution}\label{sec:appendix}

In this Appendix we discuss how the merger rate as a function of redshift is found for the case of an initial Poisson distribution of equal mass black holes. We will be particularly interested in large redshift mergers, a case that to our knowledge has not been much considered in the literature.

In a Poissonian distribution, the probability of finding a binary separated by a comoving distance $x$ and their nearest neighbour at distance $y$ is given by (see e.g. \cite{Nakamura:1997sm}) 
\be\label{eq:mergingrate_1}
Q(x,y)=16\pi^2x^2y^2n^2\exp\left[-4\pi \int_{R_\text{BH}}^y n z^2\dd z\right]\Theta(y-x) \ ,
\ee
where $n$ is the comoving number density of BHs.

There are two effects that might play an important role in the case of early time mergers, in particular for those happening before matter radiation equality.  The first one, as correctly stated in \cite{Deng:2021gkx}, has to do with with the fact that the so-called Peters time used to relate $(x,y)$ with $t$ does not provide an accurate description for very eccentric mergers.

This approximations tells us that the time to collapse from the moment of decoupling is given by \cite{Peters:1964zz}
\be\label{eq:peters}
t_{\rm p}=\frac{3}{170}\frac{r_x^4}{M^3}j^7 \quad \textrm{with} \quad j=(x/y)^3 \ ,
\ee
where $r_x = a(t_{\rm dec}) x$ is the semi-major axis of the binary at the moment of the decoupling of the binary from the cosmic flow \footnote{We set $a(t_{\rm eq}) = 1$.}, and $M$ is the mass of the black holes. While this is a good description for binaries with small eccentricities, it fails for very eccentric orbits. A binary that decouples at time $t_{\rm dec}$ and has a very large eccentricity might merge in `no time' according to Peters formula. Of course this is incorrect, since the time of merger is at least equal to the free falling time, $t_{\rm ff}$, which is given by
\be 
t_{\rm ff} = \left(\dfrac{r_x^3}{M}\right)^{\frac12} \eqdot
\ee

Therefore, Peters approximation will fail for configurations having $t_{\rm p} < \tff$. Taking into account this consideration we can write the time $\tobs$ in which a binary merges as
\be\label{eq:timeobservation}
\tobs=\tdec(x)+t_{\rm p}(x,y)+\tff(x) \eqdot
\ee

% and this is particularly relevant for the case of early mergers. The reason is that a binary decouples at the radiation era. This means that an any time $t<t_{\rm eq}$, a binary that decouples at $t$ and has a very large eccentricity might merge in 'no time' according to Peters formula. Of course this is incorrect, since the time of merger is at least equal to the free falling time, $t_{\rm ff}$. On the other hand, new binaries do not decouple during matter domination, and so all the binaries merging today have decoupled long ago, and then it is more likely that their merging time is dominated by the contribution coming from Peters formula. However, for very early mergers the free-fall time might be comparable to Peters time, and then their contribution might be important for determining the merger rate.

The time of decoupling $t_{\rm dec}$ is related to the comoving size of the binary by the following argument. A binary of size $r_{x} = a(t_{\rm dec}) x$ will decoupled at a time $t_{\rm dec}$ given that the energy density created by a binary at such distance is larger than the background energy density,
\be
\frac{M}{r_x(t_{\rm dec})^3}=\rho_{\rm back}(t_{\rm dec}) \eqdot
\ee
In order for this to be satisfied, the decoupling has to happen in radiation dominated era, where $\rho \sim \rho_{\rm eq} a^{-4}(t)$. So this implies that
\be\label{eq:dec_condition}
x = f^{\frac13}\bar{x} \ a^{\frac13}(t_{\rm dec}) \eqcom
\ee
where $\bar{x}$ is the mean comoving distance of two PBHs, defined by
\be
\bar{x}  = f^{-\frac13}\left(\dfrac{M}{\rho_{\rm eq}}\right)^{\frac13} \eqdot
\ee
Then $t_{\rm dec}$ is related to the comoving distance $x$ via Eq. \eqref{eq:dec_condition}. This relation inherits an important consideration, and has to do with the maximum size of a binary that participates in mergers at $t_{\rm obs}$. From Eq. \eqref{eq:dec_condition}, one can see that this maximum size will be given by the latest time a binary can decouple. Since binaries can only decouple in radiation dominated era, for systems merging at $\tobs>t_{\rm eq}+\tff$, the maximum size a binary can have is a constant given by 
\be\label{eq:xM_old}
\xM = f^{1/3}\bar{x} \ , \ \tobs > \teq + \tff
\ee
For mergers happening before that time, note that the latest time it can decouple and merge at time $\tobs$ is $t_{\rm dec}^{\rm max} = \tobs- \tff$. After some calculations it can be found that 
\be\label{eq:xM_new}
\xM(\tobs) \approx f^{1/3} \bar{x} \left[\tobs^2 \rho_{\rm eq}\right]^{1/12} \ , \ \tobs < \teq + \tff
\ee
which is explicitly time dependent. This time dependence has large consequences for the early merging rate.

The last piece that we need to determine is $\xm$, the minimum comoving distance participating in a merger at time $\tobs$. This is given by the distance such that a circular merger (the configuration having the largest merging time) with such semi-axis merges at time $\tobs$. This implies that all binaries separated by a distance $x<\xm$ would have merged in a time $t<\tobs$, for all the possible eccentricities. This is approximately given by
\be\label{eq:xm}
\xm = \left(\frac{170 M^3 t_{\rm obs}}{3 \left(f^{1/3}\bar{x}\right)^{4} } \right)^{1/16}f^{1/3}\bar{x} \ .
\ee

We can then perform the integral \eqref{eq:mergerrate_2}, considering $\xm$ as given in \eqref{eq:xm} and $\xM$ as given by either \eqref{eq:xM_old} or \eqref{eq:xM_new} depending on whether the observed time is after o before matter radiation equality. The function $y(x,\tobs)$ in Eq. \eqref{eq:mergerrate_2} is given by Eq. \eqref{eq:timeobservation}
\be
y(x,\tobs) = \left(\dfrac{3}{170 M^3 (f^{\frac13}\bar{x})^{12}}\right)^{1/21}\dfrac{x^{37/21}}{\left(\tobs - \tdec(x) - \tff(x)\right)^{1/21}}
\ee
Note that this function blows up at $x^* = \xM(\tobs)$, which defines the `free fall' configuration $(x,y) = (x^*,\infty)$.

Let us now consider, for a fixed observation time $\tobs$, binaries of sizes up to $x' = \lambda x^*(\tobs)$, for some $\lambda \lesssim 1$. This takes into account configurations with finite $y$, so we avoid the `free fall' configuration. The extremal configuration $(x', y(x'))$ gives us, for the Peters, free fall time and decoupling time
\be
t_p = (1 -\lambda^6)\tobs \eqcom \tff + \tdec \approx \lambda^6 \tobs
\ee
So, for $\lambda > 0.9$, the mergers will be dominated by the free fall time. In Fig. \ref{fig:high_merger_pois} we show the resulting merger rate. 
\begin{figure}[ht]
\centering
\includegraphics[width=0.7\textwidth]{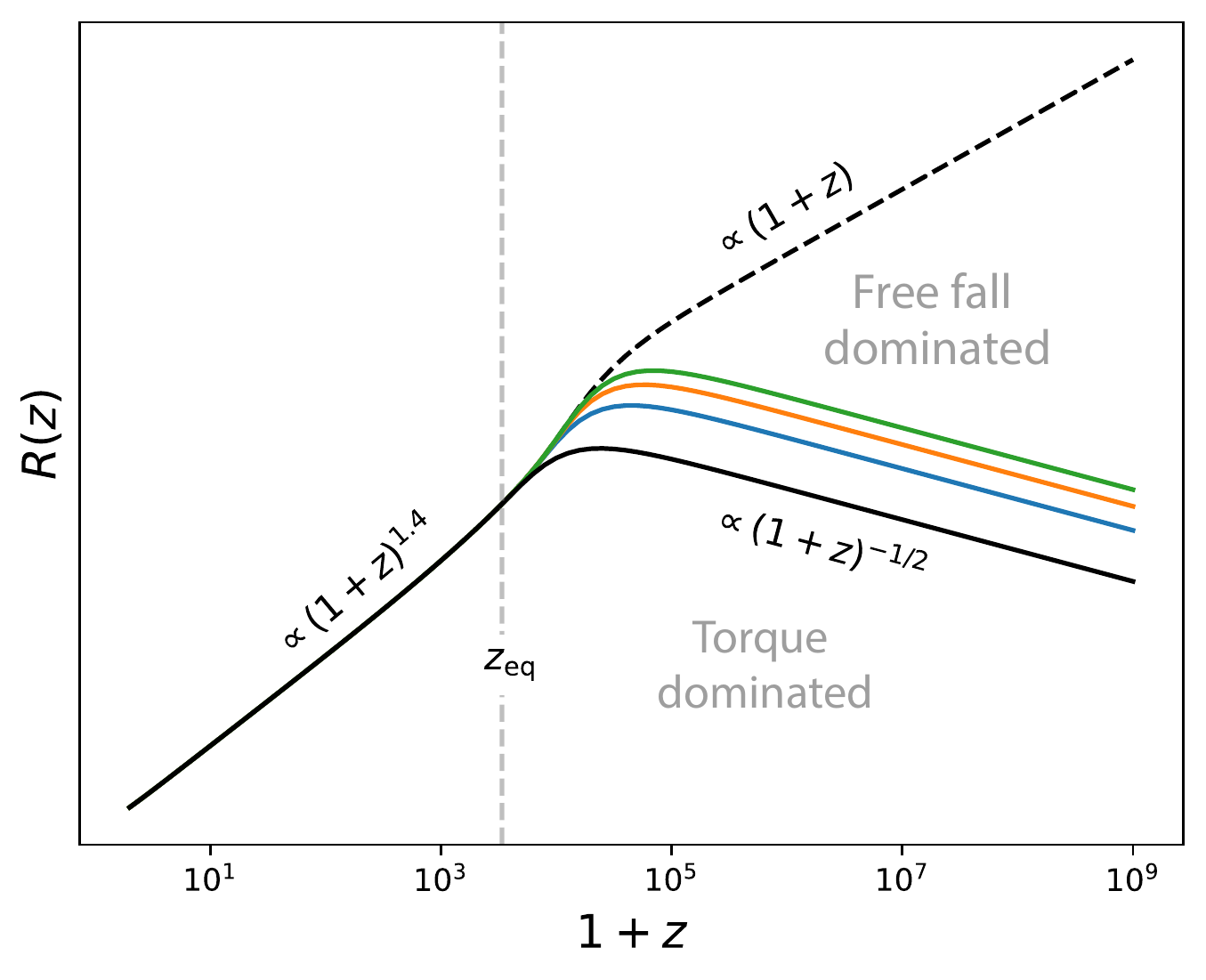}
\caption{Merger rate for a Poissonian initial distribution. Before $z=z_{\rm eq}$, the rate follows the typical scaling $\propto (1+z)^{1.4}$. After $z=z_{\rm eq}$, the merger rate increases because of the contribution of the 'free falling' orbits. However, the contribution of the mergers dominated by $t_{\rm p}$ makes the rate drop because of the shrinking of $\xM$. The dashed black line corresponds to the merger rate accounting for binaries of all sizes. The solid lines correspond to merger rates accounting for binaries of size up to $\lambda x^*$, for $\lambda = 0.9, 0.99, 0.999$ and $0.9999$ corresponding to the black, blue, orange and green colors.} \label{fig:high_merger_pois}
\end{figure}
For redshifts before matter radiation equality, the rate follows the typical scaling for the Poissionan case $R(z) \propto (1+z)^{1.4}$. At high redshift, after matter radiation equality the merger rate grows as $R(z) \propto 1+z$, mostly because of the contribution of `free fall' orbits. However, if we only consider the contribution of the mergers with torque e.g binaries of size up to $0.9x^*$, the merger rate drops as $R(z) \propto (1+z)^{-1/2}$, due to the shrinking of the maximum comoving distance participating in the merger. As we can see in Fig. \ref{fig:high_merger_pois}, this drop persists even after we consider binaries of size within $0.9x^*$ and $x^*$, although it will be visible at larger redshifts.

\newpage

\bibliographystyle{ieeetr} % We choose the &quot;plain&quot; reference style
\bibliography{sgwb.bib}

\end{document}